# Lensless speckle reconstructive spectrometer via physics-aware neural network

Junrui Liang[1, *], Min Jiang[4, *], Zhongming Huang[1], Junhong He[1], Yanting Guo[1],

Yanzhao Ke[1], Jun Ye[1,2,3], Jiangming Xu[1, *], Jun Li[1, *], Jinyong Leng[1,2,3] and Pu Zhou[1,*]

[1]*College of Advanced Interdisciplinary Studies, National University of Defense Technology, Changsha 410073, China*
[2]*Nanhu Laser Laboratory, National University of Defense Technology, Changsha 410073, China*
[3]*Hunan Provincial Key Laboratory of High Energy Laser Technology, National University of Defense Technology, Changsha 410073, China*
[4]*Test Center, National University of Defense Technology, Xi an 710106, Shaanxi, China*

\* *Co-first author*
\* *Corresponding authors: jmxu1988@163.com; lijun_gfkd@nudt.edu.cn; zhoupu203@163.com*

**Abstract:** The speckle field yielded by disordered media is extensively employed for spectral measurements. Existing speckle reconstructive spectrometers (RSs) implemented by neural networks primarily rely on supervised learning, which necessitates large-scale 'spectra-speckle' pairs. However, beyond system stability requirements for prolonged data collection, generating diverse spectra with high resolution and finely labeling them is particularly difficult. A lack of variety in datasets hinders the generalization of neural networks to new spectrum types. Here we avoid this limitation by introducing PhyspeNet, an untrained spectrum reconstruction framework combining a convolutional neural network (CNN) with a physical model of a chaotic optical cavity. Without pre-training and prior knowledge about the spectrum under test, PhyspeNet requires only a single captured speckle for various multi-wavelength reconstruction tasks. Experimentally, we demonstrate a lens-free, snapshot RS system by leveraging the one-to-many mapping between spatial and spectrum domains in a random medium. Dual-wavelength peaks separated by 2 pm can be distinguished, and a maximum working bandwidth of 40 nm is achieved with high measurement accuracy. This approach establishes a new paradigm for neural network-based RS systems, entirely eliminating reliance on datasets while ensuring that computational results exhibit a high degree of generalizability and physical explainability.

**Introduction**

Spectrometers are indispensable tools in contemporary industry and scientific research, profoundly impacting fields such as environmental sensing, biological and chemical analysis, laser characterization, and so on [1–4]. However, conventional spectrometers suffer from drawbacks, including bulky structure, upper limit on the free spectral range, and time-consuming operation resulting from scanning elements [5,6]. In recent years, speckle reconstructive spectrometers (RSs) have garnered extensive attention for their high resolution [7,8], miniaturization [9,10], unrestricted free spectral range [5,11], and fast speed [12,13]. In these systems, a beam is injected into a disordered medium, with each wavelength generating a unique intensity distribution on the sensor array, serving as a spectral "fingerprint". This mapping allows a few sensors or localized pattern profiles to cover a wide wavelength range [5], thereby fulfilling the conditions for lens-free spectral detection. The spectra cannot be read from the sensor array directly; they must be reconstructed by algorithms, such as truncated singular value decomposition (TSVD) [14,15], Lasso regression [16,17], Tikhonov regularization [18,19], and their combinations [20,21]. However, the presence of experimental noise turns spectral decoding into an ill-posed inverse problem [22]. To enhance spectral reconstruction accuracy, artificially designed regularization terms should be integrated with the objective function to be optimized [20,23]. This typically necessitates prior knowledge or assumption of the measured spectrum (e.g., sparsity) [24,25], which is impractical in novel scenarios where light sources and phenomena are never seen.

With the development of artificial intelligence (AI), end-to-end neural networks have been employed to directly learn the mapping between incident spectrum and scattered speckle images [26–28]. Such supervised learning models require high-quality mass training datasets. Because the single-wavelength dataset with high resolution is easily accessible, most current AI-empowered spectrum measurement studies focus on monochromatic light [29–34]. As for polychromatic light, generating diverse spectra with high resolving power is a significant challenge, since devices capable of precisely shaping multi-wavelength components are limited [35,36]. An alternative approach to allocate multi-wavelength is through exposure time integration; however, this method encounters intensity saturation when the bandwidth is broadened [26]. Additionally, if the spectrum generator does not bring along reference values, high-resolution spectra label need to be measured by cutting-edge elements, such as echelle gratings [35], virtual image phase arrays [36], and so on [37], making this labeling scheme resource intensive and highly costly.

Due to the dependence on substantial labeled spectra, supervised learning strategies can only yield good results on test data that share a similar distribution with the training set [38,39], while their external generalization to different spectrum types was limited. Transfer learning can mitigate this problem to some extent [7], which requires the fine-tuning of pre-trained model on a subset of new test data. However, supervised transfer learning learns features with narrow range of expected data distributions, such as specific spectrum type (e.g., only the optical frequency comb [40]), may not generalize well to wider spectra not within expectations. Moreover, transfer learning still demands additional effort and time to collect data from new test distributions and fine-tune the trained models, posing practical challenges across various applications.

In addition, pure data-driven networks lack the constraints imposed by real-world physical models, resulting in poor explainability and reliability of predictions, which is especially undesirable in spectrometry. A recent study [40] introduced a physical model within a whispering gallery mode resonator-based RS; however, the network still required many hours of training process, and data-driven loss continued to play an important role, thus rendering the limited measurement capacity of no more than four wavelengths. Moreover, this scheme was complex, involving two reconstruction stages and distinct mechanisms. In the first stage, the observed intensity must be transformed into an estimated spectrum by pseudo-inverse operation. A neural network then retrieved a high-accuracy spectrum from this non-ideal estimated one. The network in this context serves primarily to refine the results obtained by a traditional approach.

Here we demonstrate an untrained spectrum reconstruction framework built by a convolutional neural network (CNN) and the spectral-to-spatial physical model of a chaotic optical cavity. We term the framework as PhyspeNet. It can be competent for various types of multi-wavelength recovery tasks without pre-training, thus breaking the dependency on a vast amount of training data. When it comes to measuring, simply feeding a single captured speckle into PhyspeNet is sufficient. The network's weights and biases are then optimized through the interaction between the CNN and the physical model, culminating in a viable solution that adheres to the physical constraints. Furthermore, since PhyspeNet leverages the network to provide implicit priors, it does not require assumptions about the light under test or the laborious regularization term designing. We employed a compact integrating sphere (IS) as a scattering media to construct a lens-free, snapshot spectral measurement system. Using PhyspeNet and our developed speckle RS, dual-wavelength peaks divided by 2 pm can be distinguished, and a maximum working bandwidth of 40 nm is achieved with high accuracy.

**Principles**

Experimental setup

Figure 1a depicts the experimental configuration of our proposed speckle RS. Before utilizing the spectrometer, it is essential to calibrate the relationship between wavelength and spatial speckle to establish a transmission matrix of the scattering medium. At this stage, a tunable laser beam entered a fiber-coupled IS with varying wavelengths. ISs are robust optical devices with chaotic light ray behavior [5], commonly utilized in high-resolution wavelength measurements [8,41,42]. As for polychromatic light detection, we have proved that IS can offer a high ratio of independent spatial channels [43]. The internal surface material of the IS used here was barium sulfate ($BaSO_4$), selected for its high reflectance (> 95%), excellent chemical stability, and spectral neutrality. Given that IS a highly appropriate medium, we adopt it here. Novel scattering media are emerging, but it is not the scope of this study.

A fiber polarizer is employed to maintain the polarization state of the input light, ensuring that a specific wavelength consistently produce the identical speckle pattern. Light escapes the IS from a 3-mm diameter output port, and a camera captures the emerging speckle pattern. The distance between the sphere and the camera is set at 5 cm. In our experimental setup, we have effectively utilized the disorder-induced one-to-many mapping between incident spectral information and output spatial speckles. Specifically, when light undergoes multiple scattering in the IS, the input spectrum becomes thoroughly scrambled in the spatial domain, with the intensity at any given speckle location being influenced by all incoming spectral components. This observation led us to investigate the feasibility of employing directly diffracted local speckles for spectral reconstruction. Consequently, our entire spectral measurement system operates without focusing optics, significantly alleviating the stringent requirements for manual alignment.

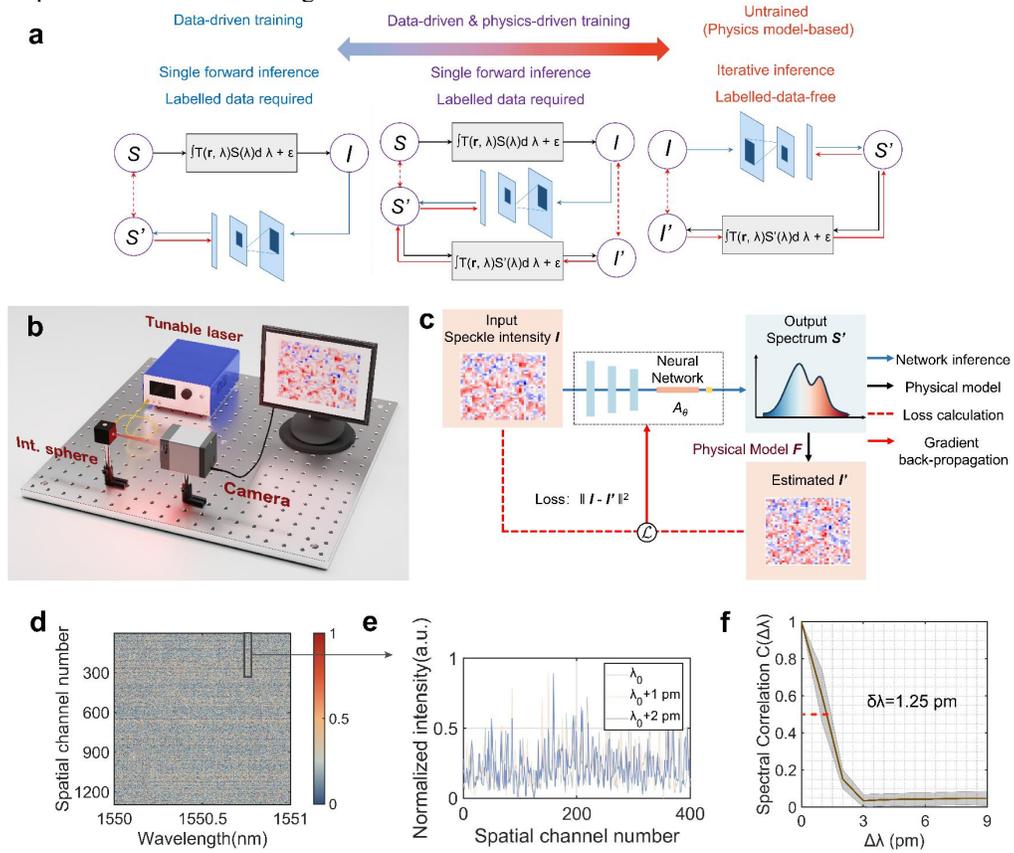

**Fig. 1. Experimental setup, PhyspeNet architecture, and forward physical model analysis.**

## Spectral reconstruction method

Before delving into the spectrum reconstruction approach employed in this paper, it is essential to introduce the working principles of typical data-driven neural networks. A vast array of labeled data $(S_k, I_k)$, $k = 1,2, ..., K$, is collected to form a dataset $D = \{(S_k, I_k), k = 1, 2, ..., K\}$, and the neural network learns the mapping relationship $A$ between these data pairs by solving the following:

$$A_{\theta^*} = \arg\min_{\theta \in \Theta} \|A_\theta(I_k) - S_k\|_2^2, \qquad \forall (S_k, I_k) \in D \qquad (1)$$

where $A_\theta$ represents the mapping function of the neural network, characterized by a set of weights and biases $\theta$ belonging to the parameter space $\Theta$, $\| \ \|_2$ represents the L2-norm. Once the neural network training is complete, the corresponding spectrum of a newly observed speckle pattern can be mapped and obtained by $S' = A_{\theta^*}(I)$. In RSs, the size $K$ of the training set can range from a few thousand to tens of thousands. Gathering such an extensive collection of speckle patterns $I_k$ and their corresponding ground-truth spectra $S_k$ experimentally is time-intensive and demands mechanical and environmental stability throughout the prolonged data acquisition process. In particular, it is challenging to generate and measure spectral data with pm-level resolution that is arbitrarily flexible. Although it is feasible to construct a training set through numerical simulations, the mapping function learned tends to perform optimally only on test images that closely resemble those in the training set. Consequently, effective generalization is limited to spectra that share the same priors used during the training phase.

Physical models can be introduced for constraints in a purely data-driven context further to improve the generalizability of the spectral reconstruction results. Here, we call it a data and physical double-driven network (DDNet). At this time, the mapping relationship $A$ can be obtained by solving the following:

$$A_{\theta^*} = \arg\min_{\theta \in \Theta} \|A_\theta(I_k) - S_k\|_2^2 + \|F(A_\theta(I_k)) - I_k\|_2^2, \qquad \forall (S_k, I_k) \in D \qquad (2)$$

where $F(\cdot)$ is defined through the forward physical model bridging the input spectra and speckle pattern. The transmission matrix is widely used for analyzing disordered media and characterizing various physical transformations. It was selected as the physical model in the previous RS and is also utilized in our work here.

By utilizing the physical model, the estimated speckle can be calculated from the spectrum reconstructed by DDNet. The difference between estimated and actual input speckles is incorporated into the objective function to guide DDNet training and prevent it from over-fitting. From Eq. (2), it can be seen that DDNet still requires a large amount of paired spectral and speckle data to be collected. The big dataset will result in a long training time for the network. In this study, a PhyspeNet model was constructed to overcome the above challenges, and its principle of spectral reconstruction can be expressed as:

$$A_{\theta^*} = \arg\min_{\theta \in \Theta} \|F(A_\theta(I)) - I\|_2^2 \qquad (3)$$

The ground-truth spectrum $S$ is notably absent from the objective functions (1) and (2), indicating that PhyspeNet does not necessitate the ground-truth spectra during training. This capability is especially valuable in high-resolution RSs, as acquiring high-resolution (picometer-level or even finer) reference data by the current commercial devices is usually challenging. As for the PhyspeNet, the interaction between $F$ and $A_\theta$ enables the prior of $I$ to be encapsulated by the neural network. Upon completion of the optimization process, the derived mapping function $A_\theta$ can subsequently be employed to reconstruct the spectra: $S' = A_{\theta^*}(I)$ Since there is no limitation on the network architecture, the proposed PhyspeNet is based on an elegant CNN, sharing the same architecture with other networks benchmarked below.

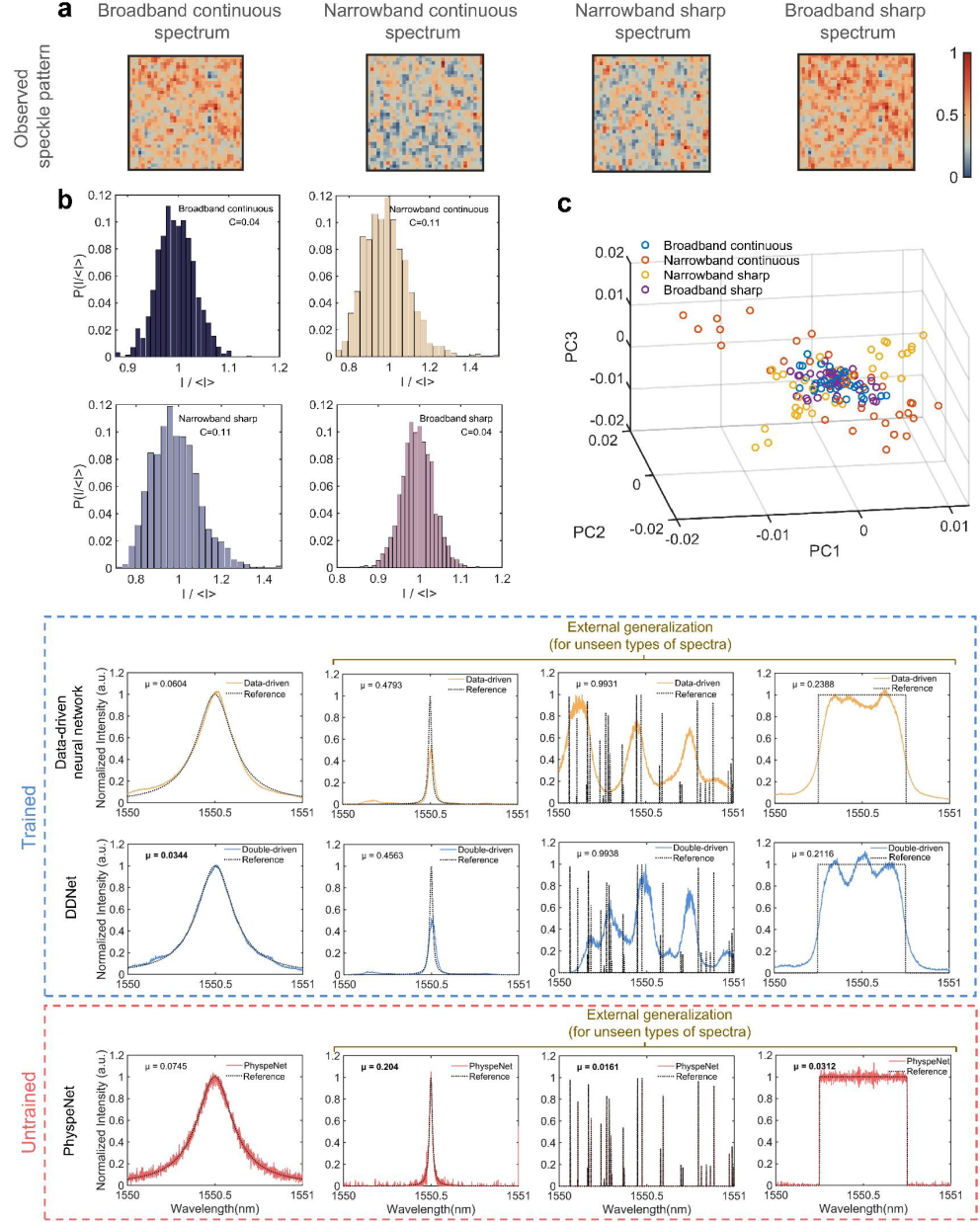

**Fig. 2. Analysis of the spectrum-dependent speckles and comparison with existing neural network models need training for spectrum recovery.**

## Results

We first compare the generalization performance of the PhyspeNet model against other spectrum recovery models that required training by using spectra with distinct features (I. broadband continuous, II. narrowband continuous, III. narrowband sharp, and IV. broadband sharp). Among these types, Spectral Type I comprises the training set for pure data-driven neural network and DDNet (see details in the Supplementary Materials), as it can be readily produced by off-the-shelf products, such as LEDs [28,44]. In contrast to Spectral Type I, the large-scale preparation of the latter three spectrum types necessitates specialized equipment, including devices with narrowband or sharp filtering edges and tunable capabilities. During the model test, we incorporate all types of spectra as detection targets. Consequently, our testing protocol mimics a real scene where a network trained on a single spectral distribution is employed to identify unknown spectra from various distributions. In particular, a Lorentz spectrum with full width at half maxima (FWHM) of 0.25 nm, a Lorentz spectrum with FWHM of 0.025 nm, a random multi-line spectrum with relative sparsity ratio (RSR) of 3%, and a rectangle spectrum with a width of 0.5 nm is selected for illustration here. Notably, the 0.25 nm FWHM Lorentz spectrum for the test here is absent from the training dataset; instead, it featured a data distribution analogous to that of Spectral Type I.

Figure 2a presents the observed speckle patterns associated with four representative spectra. The narrowband spectra depicted in the second and third columns are characterized by speckles with diminished intensity, heightened contrast, and a richer array of spatial features compared to the broadband spectra in the first and fourth columns. This phenomenon arises because the $X$ simultaneous wavelengths add in intensity [26,45], leading to a contrast reduction of $\sim X^{1/2}$ times the speckle's intensity for a single wavelength [25,46]. Consequently, some spatial features vanish or become homogenized, leading to the shift of the image domain. The intensity probability density function (PDF) distributions of the spectrum-dependent speckle are depicted in Fig. 2b. The intensity PDFs of narrowband speckles exhibit a distribution that closely aligns with a negative exponential envelope, a characteristic that is particularly pronounced in images with high contrast. Broadband speckle intensity distributions tend to be more uniform, with PDFs adhering closely to a Gaussian envelope. This alignment with the Gaussian distribution is anticipated based on the central limit theorem [47]. In Fig. 2c, we further visualize the distribution of four distinct spectral speckles in the feature space derived from Principal Component Analysis (PCA). We clearly distinguish between the speckle patterns associated with broadband spectra (blue and purple circles) and those corresponding to narrowband spectra (red and yellow circles). However, the separation of speckle patterns related to continuous and sharp spectra varies. As for broadband speckles, the features of continuous (blue circles) and sharp (purple circles) spectra are not significantly distinguishable. On the contrary, for narrowband speckles, there is a distinct separation between continuous (red circles) and sharp (yellow circles) spectra.

Figure 2c-e display the reconstruction results for various spectra using a pure data-driven neural network (yellow lines), DDNet (blue lines), and PhyspeNet (red lines). In this study, we employ the L2-norm relative error $\mu$ to assess the accuracy of spectrum reconstruction, a metric widely recognized and adopted in this field [1,21]. The error is calculated as follows:

$$\mu = \frac{\|S - S'\|_2}{\|S\|_2} \tag{4}$$

where $S'$ is the reconstructed spectrum and $S$ is the reference. Particularly, an L2-norm relative error below 0.1 is typically regarded as a stringent criterion that signifies high accuracy in spectral reconstruction [2,17,21,48]. As depicted in the first column of Fig. 2, the reconstruction results from all three models have met this specified standard, indicating a satisfactory level of internal generalization to new samples of the same type as those in the training dataset. The spectral reconstruction accuracy of the trained networks is marginally higher, with DDNet scoring the lowest relative error values. This superior performance can be attributed to incorporating additional physical prior constraints within DDNet. When blindly tested on unseen Spectral Type II, III, and IV, PhyspeNet achieved better external generalization than trained models using the same architecture. For Spectral Type II, the trained networks effectively identify the central peak region; however, the recovered intensity values are inaccurate. For Spectral Type III, the trained networks completely fail to reconstruct the narrowband multi-lines. Instead, the outcomes align with the shape of the broadband continua. For Spectral Type IV, the pure data-driven approach and DDNet can largely recover the intensity and position of the non-zero spectral components. However, these methods struggle to infer the steep cut-off edge, tending to derive the rising and falling edges of the gradient. Their relative reconstruction errors are 0.2388 and 0.2316, respectively. In stark contrast, PhyspeNet achieves a significantly lower relative error of 0.0312, demonstrating approximately seven-fold greater accuracy than the other trained networks. Furthermore, we conducted additional spectral reconstruction experiments to validate this superiority (see Supporting Materials), including Spectra I with changing FWHMs, Spectra II with shifts in center wavelength position, Spectra III with variations in RSRs, and Spectra IV with alterations in bandwidth. In summary, our results reveal that the untrained PhyspeNet model showed high reconstruction fidelity for both internal and external generalization.

In addition to its generalization advantage, another benefit of PhyspeNet lies in its utilization of the implicit prior provided by the neural network [49]. This feature eliminates the need for a data prior during the reconstruction process. In this section, we benchmark several popular spectral reconstruction algorithms, including Lasso regularization, Tikhonov regularization, and TSVD. Figure 3a illustrates that when these classical methods are used to reconstruct different spectra (without loss of generality, we have chosen Lorentzian spectra with varying FWHM), the selection of parameters is case-sensitive to achieve optimal reconstruction accuracy. As the FWHM increases, the superposition of experimental noise leads to optimal reconstruction effects occurring at higher singular value cut-off thresholds and larger L2 regularization parameters (see the first and third columns in Fig. 3a). Conversely, an increase in non-zero wavelength components requires a lesser degree of L1 regularization penalty (see the second column in Fig. 3a). By contrast, the proposed PhyspeNet focuses solely on the $\|I\text{-}F(S')\|_2^2$ during its optimization and no extra parameters need to be tuned.

Figure 3b-e shows the reconstructed spectra I~IV by PhyspeNet and three representative classical algorithms. In contrast to the aforementioned neural network methods, classical approaches do not encounter issues related to domain shifts and are universally applicable across various types of spectral reconstructions. Compared to the smooth optical spectral curve obtained through neural network inference (see Fig. 2), the spectra reconstructed using traditional methods exhibit significant artifacts. Our proposed PhyspeNet retains the universality of physics-based retrieval while leveraging the powerful data-fitting capabilities of neural networks. As shown in Fig. 3b-e, the spectra recovered by PhyspeNet demonstrate the lowest relative error values. We utilized the spectral samples from the previous section, which were employed in comparing the neural network, to validate the superior accuracy of PhyspeNet further. More analyses are provided in the supporting materials. Compared to TSVD, Lasso regularization, and Tikhonov regularization, PhyspeNet reduced the average reconstruction errors for the four spectral types by 51.77%, 84.58%, and 40.14%, respectively.

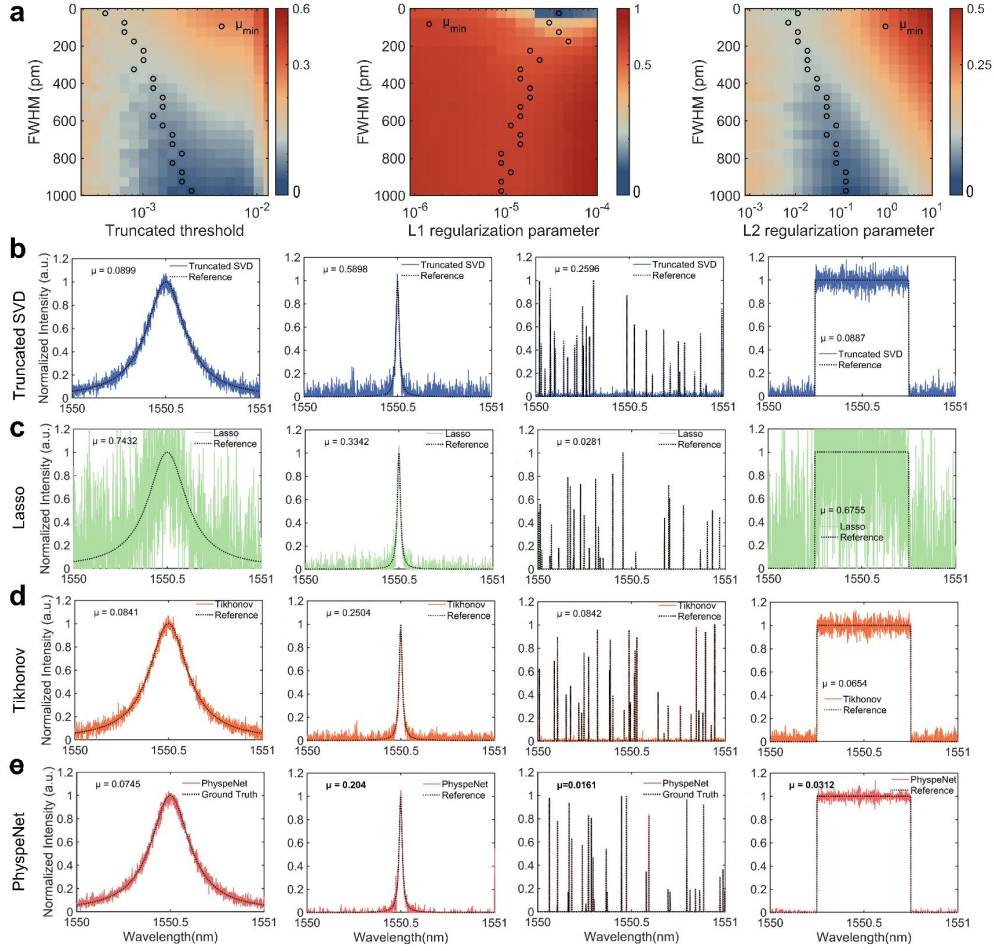

**Fig. 3. Comparison of PhyspeNet, truncated SVD, Lasso regularization, and Tikhonov regularization.**

We then applied the speckle RS based on PhyspeNet to measure various laser sources. The optical setup for testing is shown in Fig. 4a. The light under test passes through a beam splitter, with one path directed to a commercial optical spectrum analyser (OSA) for reference values, while the other path enters our RS system. As illustrated in Fig. 4b, we first measured a dichromatic beam separated by 2 pm (with central wavelengths at 1550.499 nm and 1550.501 nm, and a 3 dB linewidth of 0.04 pm). PhyspeNet could clearly distinguish between the two separate wavelengths, yielding a relative error of 0.244 and a signal-to-noise ratio (SNR) of 9.61 dB for the spectral reconstruction. The resolution capability shown in Fig. 4b aligns closely with the preliminary estimate presented in Fig. 1f. Further exploration of higher resolution is constrained by the stepping resolution of the calibration light source employed in our experiments. In the dual-peak experiment, as the two wavelengths approach one another, their spectral peaks overlap and become indistinguishable (i.e., according to the Rayleigh criterion). However, when measuring single peaks with a step of 1 pm, PhyspeNet can accurately recover their respective central wavelength positions, as illustrated in the inset of Fig. 4b. We also undertook a more challenging triple-peak detecting, where two wavelengths were positioned at 1550.199 nm and 1550.201 nm, while the third wavelength was at 1550.800 nm. The reconstruction results are displayed in Fig. 4c, demonstrating a relative error of 0.1943 and a spectral reconstruction SNR of 11.87 dB. Additionally, we measured a laser source with a linewidth of 45 pm and a fixed central wavelength of 1550.6 nm. The raw reconstruction results were down-sampled to correspond with the level of the OSA, yielding a reconstruction error of 0.5250, as shown in Fig.4c. With the increase of non-zero spectral components, the algorithm had less success, and a spurious peak occurred near the reference region. Subsequently, we evaluate the broadband operational capability of the RS system within the range of 1525 nm~1565 nm. Narrow linewidth lasers that can be tuned over a range of 40 nm were accurately reconstructed, as depicted in Fig. 4d. The average spectral reconstruction error was as low as 0.0900, with an average reconstructed SNR of 13.00 dB. The currently available calibration sources limit the broader spectral measurement range. Given that the inner surface of our IS (made of BaSO4) exhibits high reflectivity within the range of 400-2000 nm, this scheme could potentially cover both visible light and near-infrared bands if calibration and detection equipment permits it.

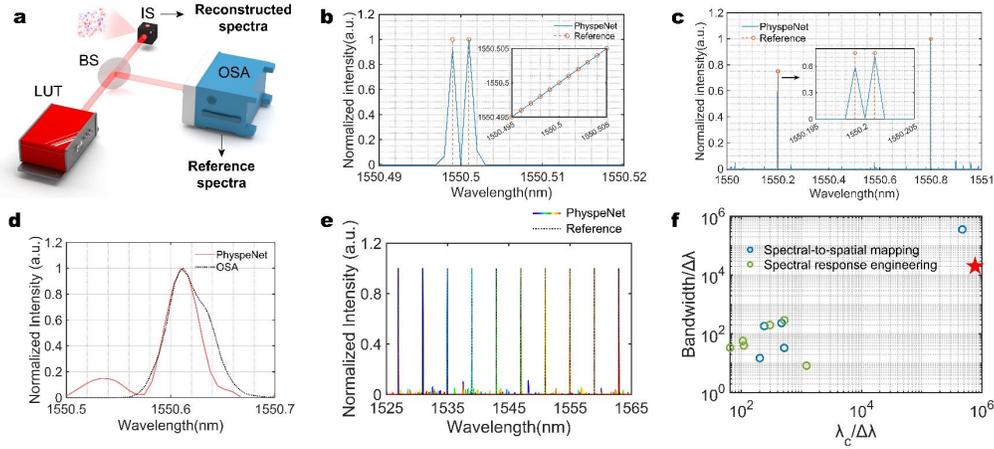

**Fig. 4.** Application of the PhyspeNet empowered speckle RS on laser source testing.